\documentclass[twocolumn,showpacs,amsmath,amssymb,showkeys,floatfix,prc,aps]{revtex4}
\usepackage{amsfonts,amsmath}
\usepackage{graphicx}
\usepackage{dcolumn} 
\usepackage{bm} 

\begin{document}

\title{The impact of hierarchy upon the values of neutrino mixing parameters}

\author{J. Escamilla-Roa$^1$, D. C. Latimer$^2$, and D. J. Ernst$^1$}

\affiliation{$^1$Department of Physics and Astronomy, Vanderbilt University,
Nashville, Tennessee 37235}

\affiliation{$^2$Department of Physics and Astronomy, University of Kentucky, Lexington, Kentucky 40506}

\date{\today}

\begin{abstract}
A neutrino-oscillation analysis is performed of the more finely binned Super-K atmospheric, MINOS, and CHOOZ data in order to examine the impact of neutrino hierarchy in this data set upon the value of 
$\theta_{13}$ and the deviation of $\theta_{23}$ from maximal mixing. 
Exact oscillation probabilities are used, thus incorporating all powers of $\theta_{13}$ and $\varepsilon :=\theta_{23}-\pi/4$. 
The extracted oscillation parameters are found to be dependent on the hierarchy, particularly for $\theta_{13}$. We find at 90\% CL are $\Delta_{32} = 2.44^{+0.26}_{-0.20}$ and $2.48^{+0.25}_{-0.22}\times 10^{-3} {\rm eV}^2$, $\varepsilon=\theta_{23}-\pi/4=0.06^{+0.06}_{-0.16}$ and $0.06^{+0.08}_{-0.17}$, and $\theta_{13}=-0.07^{+0.18}_{-0.11}$ and $-0.13^{+0.23}_{-0.16}$, for the normal and inverted hierarchy respectively. The inverted hierarchy is preferred at a statistically insignificant level of 0.3 $\sigma$. 

\end{abstract}



\maketitle

The field of neutrino oscillations has progressed rapidly over the past fifteen years. The data can largely be understood by the oscillation of the three known neutrinos \cite{Maki:1962,Pontecorvo:1968}. Oscillation phenomenology invokes a unitary matrix that relates the flavor basis, in which the neutrinos are created or destroyed, to the mass basis, in which the neutrinos propagate through vacuum.  This matrix can be written in terms of three real mixing angles, $\theta_{12}$, $\theta_{23}$, and $\theta_{13}$, and a phase $\delta$ that determines CP violation.
Oscillations also require nonzero neutrino mass differences, being dependent upon the difference of the square of the masses, $\Delta_{ij}:=m_i^2-m_j^2$ with $m_i$ the mass of neutrino $i$. A recent analysis \cite{Schwetz:2008} reports the present knowledge of the values of the oscillation parameters.  In particular, we note that only an upper limit exists on the size of the ``reactor" mixing angle $| \theta_{13} |=0.19^{+0.12}_{-0.19}$ (1$\sigma$), a constraint which arises, in part, from the null oscillation result of the CHOOZ experiment \cite{chooz}.  A nonzero value of $\theta_{13}$ is requisite for the existence of CP violation in neutrino oscillations; hence, we presently have no knowledge of the value of $\delta$.  We also have no knowledge of the ordering of the neutrino mass eigenstate; namely, is $m_3$ greater or less than $m_1$ and $m_2$?  The former (latter) situation is termed the normal (inverted) hierarchy.

Recently, hints of nonzero $\theta_{13}$ have been reported as a means by which to ease the tension between the determination of the oscillation parameters $\theta_{12}$ and $\Delta_{21}$ by the solar and KamLAND experiments \cite{Balantekin:2008,Fogli:2008}. 
At a smaller significance, analyses of atmospheric data also hint at nonzero $\theta_{13}$ \cite{Fogli:2005,Roa:2009}, though one study shows that the significance of the results are dependent upon the precise nature of the statistical analysis \cite{Maltoni:2008ka}.  Furthermore, an analysis of the updated atmospheric data finds no preference for nonzero $\theta_{13}$ \cite{Wendell:2010md}; however, this analysis employs approximate oscillation formulae and cannot be directly compared to these other works.
Should these various hints be cleanly confirmed in current neutrino oscillation experiments \cite{dchooz,dyb,reno}, then one might be able to attack the issue of CP violation in the lepton sector.  The situation, however, is confounded by the existence of various degeneracies amongst the parameters \cite{Barger:2001,Minakata:2010zn} which can render ambiguous their extraction from experiments.
To break the degeneracies, one needs to combine the results of experiments, e.g., superbeams, operating at different energies and/or baselines \cite{BurguetCastell:2002,Barger:2001,Wang:2001,Minakata:2002,Whisnant:2002,Huber:2003a,Huber:2003b,Donini:2004,Mena:2005a,Ishitsuka:2005,Gandhi:2006,Agarwalla:2006,Bernabeu:2009}. 
 In the near future, the T2K \cite{Hayato:2005} and NO$\nu$A \cite{Ayres:2005} experiments may be able  to determine the hierarchy \cite{Huber:2004,Minakata:2003,Mena:2004,Mena:2005b}.
 Atmospheric neutrinos experiments can certainly be of some use in unraveling these unknowns as  upgoing neutrinos travel through the earth over a large range of baselines and energies 
\cite{Bernabeu:2001,Indumathi:2004,Gandhi:2004, Huber:2005,Petcov:2005,
Campagne:2006,Indumathi:2006,Gandhi:2006,Gandhi:2007b,Samanta:2009qw}.
In Ref.~\cite{Roa:2009wp}, with fixed solar parameters, we analyzed the determination of the remaining oscillation parameters from the atmospheric, MINOS \cite{Minos}, CHOOZ \cite{chooz}, and K2K \cite{k2k} experiments assuming CP is conserved (i.e., taking $\delta=0,\pi$); this work, unlike others, included the more finely binned data from Super-K  \cite{Hosaka:2006}.  With that data, we now analyze the effect of hierarchy on our parameter extraction and explore any preference for a particular hierarchy by the data.

From Ref.~\cite{latimer:2004}, we note that vacuum oscillations are invariant under the map 
\begin{equation}
\begin{tabular}{lllll}
$\theta_{12} \mapsto$ & $\pi/2-\theta_{12}$ , &$\qquad$ & $\Delta_{21} \mapsto$& $\Delta_{21}$ ,   \\
$\theta_{13} \mapsto$ & $-\theta_{13}$ ,  && $\Delta_{32} \mapsto$& $-\Delta_{31}$ ,  \\
$\theta_{23} \mapsto$ & $\theta_{23}$ ,  && $\Delta_{31} \mapsto$& $-\Delta_{32}$ .
\end{tabular}  
\label{vac_sym}
\end{equation}
In the limit of vanishing $\theta_{13}$, we see that both hierarchies produce identical oscillation probabilities, provided the proper adjustment is made to $\theta_{12}$.  
Of course, matter effects \cite{Mikheyev:1985,Wolfenstein:1978} in the solar sector break the symmetry between these hierarchies requiring $\theta_{12} \le \pi/4$, provided one assumes $\Delta_{21}>0$.  
To remain consistent with the solar data, a broken symmetry is introduced where one does not implement the transformation of $\theta_{12}$ or $\theta_{13}$ in Eq.~(\ref{vac_sym}), but does  make the transformation of the mass-squared 
differences.  Given that the octant of $\theta_{12}$ is known from matter effects, it is, in principle, 
possible to distinguish hierarchy through the precision measurement of vacuum oscillation channels 
\cite{Petcov:2001sy,Nunokawa:2005nx,deGouvea:2005hk,Gandhi:2009};
however, for $\theta_{13}\sim0$, hierarchy is difficult to discern in part because of the large separation between mass-squared differences, $\Delta_{21} \ll | \Delta_{32}| $.  
If $\theta_{13}$ is sufficiently nonzero, then matter effects provide the most promising avenue by which one might determine neutrino hierarchy.

\begin{figure}
\begin{center}
\includegraphics*[width=3in]{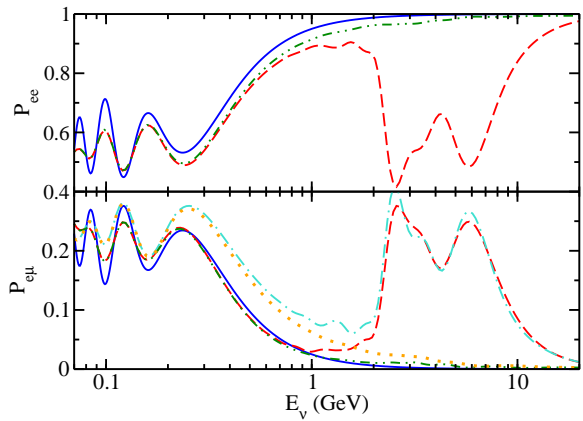}
\caption{(color online) The oscillation probabilities ${\mathcal P}_{ee}$  and ${\mathcal P}_{e\mu}$  versus neutrino energy for bin I ($-1.0<\cos\varphi<-0.8$) of the Super-K experiment using a two-density model of the earth. The (blue) solid curves are for both hierarchies and for $\theta_{13}=0$; the (red) dashed curves are for the NH and $\theta_{13}=+0.15$; the (green) dot-dot-dash curves are for the IH and $\theta_{13}=+0.15$. 
For ${\mathcal P}_{e\mu}$, the (turquoise) dot-dash curve is the NH and  $\theta_{13}=-0.15$; the (orange) dotted curve is the IH  and $\theta_{13}=-0.15$.}
\label{fig1}
\end{center}
\end{figure}

For nonzero $\theta_{13}$, resonant enhancement of the oscillation probability $\mathcal{P}_{e\mu}$ can occur over long baselines which traverse the earth's mantle and/or core for energies around 3 to 7 GeV \cite{Banuls:2001}. 
 Using a two density model of the Earth \cite{Dziewonski:1981} given by a mantle density of 4.5 g/cm$^3$ and a core density of 11.5 g/cm$^3$ with radius 3486 km, this enhancement is apparent in Fig.~\ref{fig1} which shows the $\mathcal{P}_{ee}$ and $\mathcal{P}_{e\mu}$ oscillation channels for neutrinos traveling roughly the entire diameter of the earth.  The resonance at 6 GeV (2.5 GeV) arises from the mantle (core) density.
Resonances only occur for neutrinos in the normal hierarchy (NH); however, a similar resonance occurs for anti-neutrinos in the inverted hierarchy (IH).    If $\theta_{13}$ were significantly nonzero, then one could use the presence or absence of the resonance to discern which hierarchy is realized by nature.  To do so, one would either need to have a relatively pure neutrino (or antineutrino) source or a detector, such as a magnetized iron calorimeter, which could distinguish neutrino from antineutrino \cite{Bernabeu:2001,Indumathi:2004,Gandhi:2006,Gandhi:2007b,Samanta:2009qw}.  
Unfortunately, the atmospheric neutrino and anti-neutrino spectra are roughly equivalent, and  water Cerenkov detectors, like Super-K, cannot resolve neutrino from anti-neutrino.  However, there is some sensitivity to hierarchy as the neutrino cross-section is around a factor of two greater than the anti-neutrino cross section over the relevant energy range.

In Refs.~\cite{Roa:2009,Roa:2009wp}, we found that atmospheric neutrino data had an interesting impact on the extraction of $\theta_{13}$ from the global data set.  Treating the allowed parameter range of $\theta_{13}$ as a continuous set (i.e., allowing negative values for the mixing angle is equivalent \cite{Latimer:2005a,Latimer:2005b} to setting $\delta = \pi$), it was shown that the atmospheric data placed a stringent upper bound on $\theta_{13}$.  The atmospheric data only weakly bounds $\theta_{13}$ from below with the CHOOZ data providing the dominant constraint.  As such, the data shows a statistically insignificant preference for $\theta_{13}<0$.  These constraints were traced to an excess of $e$-like events in the sub-GeV data set, a region where the oscillations due to the ``solar" mass-squared difference is no longer trivial over very long baselines.  Analytical expressions from previous studies of atmospheric neutrinos at these baselines and energies show that mass hierarchy will have at most a 10\% impact upon the effective value of $\theta_{13}$ in matter \cite{Peres:2004,Latimer:2005c}. 
This is apparent in Fig.~\ref{fig1}. Below 1 GeV, for both ${\mathcal P}_{ee}$ and ${\mathcal P}_{e\mu}$, the NH, (red) dashed curve, and IH, (green) dot-dot-dash curve, for $\theta_{13}=+0.15$ nearly overlap.  ${\mathcal P}_{ee}$ is a 
function of $\theta_{13}^2$ and thus the $\theta_{13}=-0.15$ curve is identical to the $+0.15$ curve. For ${\mathcal P}_{e\mu}$ and $\theta_{13}=-0.15$, 
the two hierarchies, the NH (turquoise) dot-dash curve and the IH (orange) dotted curve, also yield a nearly identical oscillation probability.  Hierarchy should have little impact on the asymmetric nature of the bounds on $\theta_{13}$, as this originates from the low energy data. 

From 1 GeV up to 20 GeV, we see the resonance enhancement of oscillations for the NH and nonzero $\theta_{13}$, the (red) dashed curve, $\theta_{13}=+0.15$, and the (turquoise) dot-dash curve, $\theta_{13}=-0.15$.  The resonance is absent for the IH for $\theta_{13}=+0.15$, the (green) dot-dot-dash curve, and $\theta_{13}=-0.15$, the (orange) dotted curve. 
Also, we see that the linear in $\theta_{13}$ terms are small in this region. Thus the resonances provide information about the hierarchy and the magnitude of $\theta_{13}$, but not its sign.

\begin{figure}
\begin{center}
\includegraphics*[width=3in]{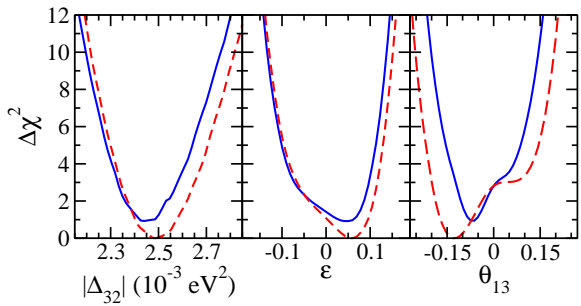}
\caption{(color online) $\Delta\chi^2$ versus the three varied parameters for an analysis that utilizes  the Super-K atmospheric, MINOS, and CHOOZ data. The solid (blue) curve is for the NH and the dashed (red) curve is for the IH.}
\label{fig2}
\end{center}
\end{figure}

Turning to the data, in Fig.~\ref{fig2}, we plot $\Delta\chi^2$ versus the three parameters we vary: $\Delta_{32}$, $\varepsilon := \theta_{23}-\pi/4$, and $\theta_{13}$.  Here, we set $\Delta\chi^2 := \chi^2-\chi^\text{2,IH}_\text{min}$ with $\chi^\text{2,IH}_\text{min}$ the minimum value of $\chi^2$ for the IH.
The analysis tools used here are described in detail in Ref.~\cite{Roa:2009wp}. This analysis
incorporates the exact oscillation probabilities,
and it makes use of the more finely-binned atmospheric data \cite{Hosaka:2006}.  We also include the MINOS \cite{Minos} and CHOOZ \cite{chooz} data.
We fix the solar parameters at their best fit values \cite{Schwetz:2008}: $\Delta_{21}  =  7.65\times 10^{-5}~{\rm eV}^2$ and $\theta_{12}= 0.584$. We are able to do this
because, as can be inferred from Fig.~\ref{fig1}, the solar data is at sufficiently low energies that it makes no distinction between the hierarchies.  To obtain the $\chi^2$ curves we minimize the remaining varied parameters.

Since MINOS provides the strongest constraints on $\Delta_{32}$ and it does not have significant matter effects, we expect small differences between the hierarchies, which we find.
The minimum value of $|\Delta_{32}|$ for the IH occurs at a slightly higher value than for the NH. The resulting values at the 90\% confidence level ($\Delta\chi^2=6.25$ for a three parameter fit) for $|\Delta_{32}|$ are $|\Delta_{32}| = 2.44^{+0.26}_{-0.20}$ eV$^2$ and $2.48^{+0.25}_{-0.22}\times 10^{-3}$ eV$^2$ for the NH and IH, respectively.  The fact that $|\Delta_{32}|$ in the IH is larger than that in the NH is not surprising as suggested by the mappings between the mass-squared differences in Eq.~(\ref{vac_sym}).   The difference in $\chi^2$ between the minima for the two hierarchies is 0.9 which is a 0.3 $\sigma$ effect for a three parameter fit.

Also in Fig.~\ref{fig2}, we present $\Delta \chi^2$ versus $\varepsilon$. As we vary all three parameters, the difference in $\chi^2$ between the two minima corresponding to the two hierarchies is the same in all three panels. The best fit values are
$\varepsilon = 0.06^{+0.06}_{-0.16}$ and $0.06^{+0.08}_{-0.17}$ for the NH and IH, respectively. This indicates a preference for $\theta_{23}$ to lie in the second octant ($\varepsilon$ is positive) by $\Delta\chi^2$ of 0.6 and 1.2, or 0.2$\sigma$ and 0.4$\sigma$, for the NH and IH, respectively.

In the last panel of Fig.~\ref{fig2}, we present $\Delta \chi^2$ versus $\theta_{13}$.
 There are differences between the hierarchies in the location of the minima and in the errors for $\theta_{13}$. Here we find the best-fit values for $\theta_{13}$ are $-0.07^{+0.18}_{-0.11}$ and $-0.13^{+0.23}_{-0.16}$ for the NH and IH, respectively. 
This implies that $\theta_{13}$ is negative and non-zero by $\Delta\chi^2$ of 1.8 and 2.8, or 0.5$\sigma$ and 0.8$\sigma$, for the NH and IH, respectively. In Ref.~\cite{Latimer:2005c}, we showed that a negative value for $\theta_{13}$  allows $\theta_{23}$ to lie in the second octant while still maintaining an excess of sub-GeV $e$-like events.

These results follow what we expect from Fig.~\ref{fig1}. We find a preference for a negative $\theta_{13}$ independent of hierarchy as this arises from the excess  of sub-GeV $e$-like events seen in the data;  in this region,  the hierarchy question is not relevant.  In extracting this mixing angle from the data, the main difference between the hierarchies is that the minimum value of $\chi^2$ shifts further from zero and the error bars for the IH are larger.  The larger error bars have been previously noted \cite{Hosaka:2006,Ge:2008}. 
The origin of this and the other differences  between the hierarchies arises from the high-energy MSW resonances present for NH neutrinos and IH anti-neutrinos for nonzero $\theta_{13}$.
Due to the difference in cross sections, neutrinos have a greater impact upon  the data over anti-neutrinos.    As the atmospheric data is less sensitive to antineutrinos, a larger value of 
$|\theta_{13}|$ is needed for the IH to account for the data in the resonance region.
For the normal hierarchy which has the resonances in the dominant neutrino channel, the atmospheric data is able to reasonably bound $\theta_{13}$; however, for the IH,  bounds from the atmospheric data on $\theta_{13}$ are less stringent.

We have found that the present world's data have statistically insignificant implications for four important questions: the magnitude of $\theta_{13}$, the sign of $\theta_{13}$, the octant of $\theta_{23}$, and neutrino hierarchy. What might be required for atmospheric data to provide some significant hint of hierarchy?  The answer depends on the value of $\theta_{13}$, since this controls the size of the matter resonances. We can provide a rough estimate by looking at the difference between the hierarchies as predicted by the theory for each data bin for the fully contained atmospheric data. The upgoing neutrinos provide the greatest impact; for these data bins, we find that the theoretical difference between the hierarchy values is about one half of the present statistical error of the experimental result for $\theta_{13}=\pm 0.15$. 
If $\theta_{13}$ is in fact this large, then a reduction of the statistical error bars by a factor of two, or an increase in the total number of events by a factor of four, could begin to produce statistically significant indications of hierarchy.

We have investigated the question as to how neutrino hierarchy affects the extraction of $\theta_{13}$ from atmospheric and long baseline experiments. We have included an exact expressions for the oscillation probabilities, which necessarily contains all linear and higher order terms in $\theta_{13}$ and $\varepsilon$. We have also used the more finely-binned Super-K atmospheric data. The IH is preferred at a statistically insignificant level. The extracted value of $\theta_{13}$ differs between the two hierarchies, with $\theta_{13}=-0.07^{+0.18}_{-0.11}$ and $-0.13^{+0.23}_{-0.16}$ for the NH and IH, respectively. The error on $\theta_{13}$ is smaller in the NH as the presence of the high-energy resonances in the dominant neutrino channel leads to the Super-K atmospheric data restricting the value of $\theta_{13}$, an effect that is absent for neutrinos in the IH. The extracted values for $\theta_{23}$ and $\Delta_{32}$ are $\varepsilon = 0.06^{+0.06}_{-0.16}$ and $0.06^{+0.08}_{-0.17}$, $\Delta_{32} = 2.44^{+0.26}_{-0.20}$ and $2.48^{+0.25}_{-0.22}\times 10^{-3} {\rm eV}^2$, for the NH and IH, respectively.  

\section{ACKNOWLEDGMENTS}

The work of J.~E.~-R. is supported, in part, by CONACyT, Mexico. The work of D.~C.~L.~is supported, in part, by
US Department of Energy Grant DE-FG02-96ER40989. The work of D.~J.~E.~is supported, in part, by US Department of
Energy Grant DE-FG02-96ER40975. 

\bibliography{hier_sf.bib}

\end{document}